\documentclass[10pt]{wlscirep} 
\usepackage[utf8]{inputenc}
\usepackage[T1]{fontenc}
\usepackage{color}
\usepackage{url}

\usepackage{color}
\usepackage{graphicx}
\usepackage{tabularx}
\usepackage{todonotes}
\usepackage{hyperref}
\usepackage{amsmath}
\usepackage{xcolor}
\usepackage{comment}
\usepackage{caption}
\usepackage{subcaption}
\usepackage{multirow}
\definecolor{light-gray}{gray}{0.95}
\newcommand{\code}[1]{\colorbox{light-gray}{\texttt{#1}}}
\graphicspath{{./images/}}

\usepackage{wrapfig}

\title{The evolution of scientific literature as metastable knowledge states}

\author[1]{Sai Dileep Koneru}
\author[2]{David Rench McCauley}
\author[2]{Michael C. Smith}
\author[2]{David Guarrera}
\author[2]{Jenn Robinson}
\author[1,*]{Sarah Rajtmajer}
\affil[1]{The Pennsylvania State University, University Park, PA, USA}
\affil[2]{Ernst \& Young, McLean, VA, USA}
\affil[*]{smr48@psu.edu}
\usepackage{titlesec}
\titleformat{\section}
  [hang]
  {\normalfont\bfseries\Large}
  {}
  {0pt}
  {}
\renewcommand{\thesection}{}
\renewcommand{\thesubsection}{\arabic{subsection}}

\keywords{metascience, bibliometrics, scientometrics, network science, topic modeling, interdisciplinarity, information diffusion}

\begin{abstract}
The problem of identifying common concepts in the sciences and deciding when new ideas have emerged is an open one. Metascience researchers have sought to formalize principles underlying stages in the life-cycle of scientific research, determine how knowledge is transferred between scientists and stakeholders, and understand how new ideas are generated and take hold.  
Here, we model the state of scientific knowledge immediately preceding new directions of research as a metastable state and the creation of new concepts as combinatorial innovation. We find that, through the combined use of natural language clustering and citation graph analysis, we can predict the evolution of ideas over time and thus connect a single scientific article to past and future concepts in a way that goes beyond traditional citation and reference connections.  
\end{abstract}
\begin{document}

\maketitle

\vspace{-0.9cm}
\section*{Introduction}

Early work in metascience can be traced back at least half a century, \cite{morris1946significance} although it has been only in the last decade or so that a robust literature has been seeded exploring co-authorship networks, citation networks, topical networks and similar static and one-dimensional representations of complex interactions amongst researchers and their work. Much of this has been powered by the increased availability of digital data on scientific processes, improvements in information retrieval, network science, machine learning, and computational power, allowing researchers to derive meaningful insights.
A substantial subset of this literature has focused on quantifying and predicting success in publishing -- how we should measure success, who will have it, and what factors contribute to having it. Seminal work has focused on modeling citation patterns for papers \cite{wang2013quantifying} and researchers \cite{sinatra2016quantifying}, with more recent work setting out to explain hot streaks in researchers' career trajectories\cite{liu2018hot}, unique patterns of productivity and collaboration amongst the scientific elite\cite{li2020scientific}, and even the role of luck in driving scientific success\cite{pluchino2019exploring,janosov2020success}.
We are also seeing the emergence of metascience as a social movement \cite{peterson2020metascience}, catalyzed by the last decade's reproducibility crisis\cite{schooler2014metascience}, aiming to describe and evaluate science at a macro scale in order to diagnose biases in research practice\cite{lariviere2013bibliometrics,hofstra2020diversity}, highlight flaws in publication processes\cite{franco2014publication}, understand how researchers select new work to pursue\cite{rzhetsky2015choosing,jia2017quantifying}, identify opportunities for increased efficiency (e.g., automated hypothesis generation \cite{spangler2014automated}), and forecast emergence of research topics \cite{prabhakaran2016predicting,chen2018modeling}. 

Prior work on the evolution of research can be broadly viewed in three categories based on method: network-based, language-based, and hybrid methods using both networks and language. Language-based methods commonly use topic models such as Latent Dirichlet Allocation (LDA) and predict changes in topics\cite{kleminski2017identifying,uban2021studying}. 
Other language-based approaches include tracking usage of keywords\cite{faust2018documenting}, analyzing linguistic context\cite{prabhakaran2016predicting}, and modeling topics sequentially\cite{chen2018modeling}. Studies using network-based methods usually 
use citation networks and community detection algorithms such as topological clustering methods \cite{shibata2008detecting} or clique percolation methods\cite{salatino2018augur} to identify emergence of new fields, while other network approaches include usage of temporal\cite{sun2020evolution}, multiplex\cite{zamani2020evolution} networks, projections of citation networks such as 
co-authorship\cite{sarigol2014predicting,sun2016mapping}.
Hybrid usage of both language- and network-based methods to predict the evolution of scientific fields includes keyword-generated networks used to predict changes in topics \cite{krenn2020predicting} or approaches that mostly rely on network analysis, applying linguistic techniques such as LDA for explanatory labels only\cite{sasaki2020emerging}. Still others \cite{zhang2017detecting} have used LDA and co-occurrence networks of topics to study changes in knowledge-based systems. However, to the best of our knowledge, these hybrid methods do not incorporate state-of-the-art language embeddings, nor do they incorporate insights from both the language models and the citation network. It is our hypothesis that the only way to truly capture the evolution of ideas and knowledge in the literature is through the integration of network and linguistic techniques. 

A premise of the work discussed in this paper is that neither citation networks alone (or derivatives thereof) nor purely language-driven models of the scientific corpus can explain the evolution of fields and the emergence of new ideas. We show that these two frameworks capture overlapping but distinct and complementary aspects of dynamics in scientific research. 
We use pre-trained neural network models \cite{cohan2020specter} to generate vectorized representations of the literature while separately leveraging citation network measures (e.g., betweenness centrality), combining these two inputs to build predictive models of topical evolution. 
The intuition behind the mechanisms explored herein is that scientific disciplines can be described at a high level by aggregation of related ideas. When a discipline is beginning to show signs of fracture or change via the emergence or synthesis of new ideas, we model this moment borrowing from physics the concept of \emph{metastability}: a state 
easily perturbed into a new state. We suggest that measures of interdisciplinarity may be indicators of this transition and thus useful predictors of change in the scientific ecosystem. 

Recent efforts have elevated the role of interdisciplinarity in scientific practice\cite{klein1990interdisciplinarity,jacobs2009interdisciplinarity,repko2019introduction,pan2012evolution}. Prior work has shown interdisciplinarity to have an effect on innovation and research impact\cite{molas2014relationship,hofstra2020diversity}. Calls for collaboration across disciplines are prominent throughout research institutions and funding agencies\footnote{See, e.g., the U.S. National Science Foundation's Growing Convergence Research program: \url{https://www.nsf.gov/od/oia/growing-convergence-research/index.jsp}.} but some have argued that the promises of interdisciplinarity are overstated and misplaced\cite{jacobs2014defense}. The bibliometric community has offered a data-driven framing for interdisciplinary studies, e.g., defining interdisciplinarity as a process of integrating different bodies of knowledge\cite{wagner2011approaches,porter2009science}.

The definition of interdisciplinarity 
varies broadly in the literature, with different definitions capturing different aspects of this concept\cite{wang2020consistency}, and can be broadly classified into two groups: subject-based and network-based definitions\cite{wang2020consistency}. 
Subject-based metrics rely on multi-classification systems to calculate interdisciplinarity, leaning on pre-defined subject categories, e.g., from the Web of Science (WoS)\cite{WoS}. 
These approaches generally are imposed at the journal level, focusing on the distribution of subject categories, e.g., percentage of references cited by publications in journals outside a journal of interest's category\cite{porter1985indicator,morillo2001approach}. In some cases, metrics are borrowed from other fields, such as the Gini index from economics and Shannon entropy from information theory, to quantify diversity\cite{wang2015interdisciplinarity}); these are also based on pre-made categories. 
Alternatively, network-based interdisciplinarity metrics are often assessed based on the location of a publication in a citation network\cite{leydesdorff2007betweenness}, with centrality measures frequently being the focus. For example, betweenness centrality
, which is independent of third-party categorization, was one of the first metrics used in this way\cite{leydesdorff2007betweenness,leydesdorff2011indicators} and has likewise been used to predict future network trends\cite{gao2021potential,chen2009towards}. 

\begin{figure}[h!]
\vspace{-0.2cm}
\centering
\includegraphics[width=\textwidth]{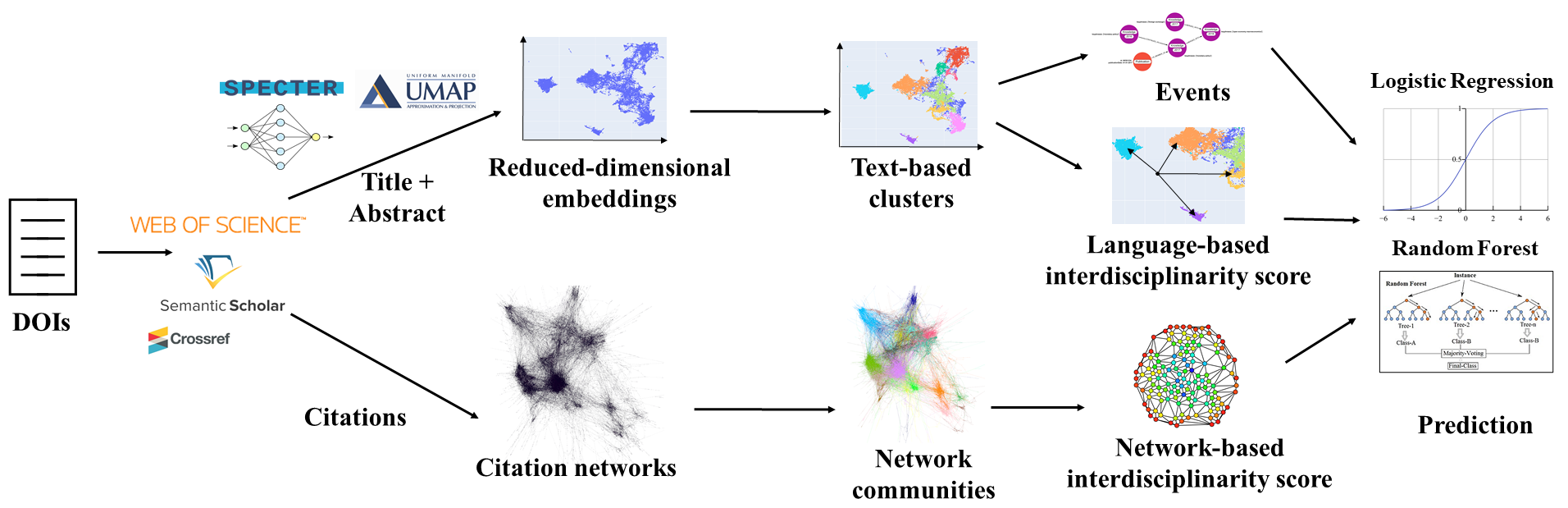}
\caption{\textbf{Data analysis workflows.} (Top) \textbf{Text-based analysis.} Title and abstract are concatenated and input to a language embedding model, then dimensionally reduced and fed into a clustering algorithm; clusters of embedded papers are then used for event modeling and interdisciplinarity scoring. (Bottom) \textbf{Citation-based analysis.} Citation information is used to create undirected citation graphs; the Louvain algorithm is used to identify network communities and betweenness centrality is used for interdisciplinarity scoring. Interdisciplinary metrics are jointly used to predict disciplinary evolution.}
\label{fig:workflow}
\vspace{-0.2cm}
\end{figure}

To study knowledge evolution in the scientific literature, we: (1) develop methods that utilize transformers-based language models and unsupervised clustering to track the evolution  of ideas over time; (2) quantify interdisciplinarity using complementary text- and citation-based metrics; and (3) explore the utility of metastability, measured through interdisciplinarity, as a predictor of scientific evolutionary events (Fig. \ref{fig:workflow}).

\vspace{-0.2cm}
\section*{Dataset}

Our dataset contains detailed records of 19,177 scientific papers published in the years 2011 through 2018,  with 2300 to 2500 papers for each year, representing a substantial stratified random sample of papers published in 62 prominent journals from the following disciplines as strata: Criminology; Economics and Finance; Education; Health; Management; Marketing and Organizational Behavior; Political Science; Psychology; Public Administration; and Sociology.\footnote{The dataset was collected in conjunction with DARPA's SCORE program. For a complete listing of journals see\cite{alipourfard2021systematizing}.}
Metadata for these were collected using the Web of Science as a primary source. Digital Object Identifiers (DOIs) were used to merge WoS records with Semantic Scholar (S2) records\cite{ammar2018construction,lo2019s2orc} for completeness of metadata coverage and author name disambiguation. When DOIs were not available from WoS, we used Crossref \cite{lammey2015crossref} to fill in missing DOIs for more complete record linking between the Web of Science and Semantic Scholar. 
For citation network analyses, we also included all papers referenced by these papers. 
Our complete dataset includes records of 839,096 papers and about 1.45 million citations. 

\section*{Methods}

We use parallel workflows to model dynamics in bibliometric data -- one based on text and one based on citation networks (Fig. \ref{fig:workflow}). 
For each we derive a measure of interdisciplinarity useful for prediction of knowledge evolution and we describe our explanatory and predictive experiments to evaluate our measures.

\vspace{-0.1cm}
\subsection*{SPECTER-based topic modeling}
We use language-embedding-based topic modeling to identify topics within our corpus for a given year. To do so, we extract embeddings for each publication in our dataset using the concatenated title and abstract as an input to SPECTER (Scientific Paper Embeddings using Citation-informed TransformERs)\cite{cohan2020specter}, a 
model for generating document-level embeddings of scientific documents via pre-training on scientific papers and their citation graphs.\footnote{Specifically, we use the \code{huggingface} implementation \cite{wolf-etal-2020-transformers} of the pre-trained SPECTER model.} SPECTER embeddings have been shown to outperform competitive baselines on benchmark document-level tasks such as citation prediction, document classification and recommendation\cite{cohan2020specter}.

To identify disciplines and subdisciplines, we use an unsupervised, non-parametric, hierarchical clustering algorithm, Hierarchical Density-Based Spatial Clustering of Applications with Noise (HDBSCAN)\cite{mcinnes2017hdbscan}. Specifically, we soft-cluster SPECTER embeddings to reflect that papers may belong to multiple (sub)disciplines with different probabilities. As the performance of HDBSCAN generally reduces as the dimensionality of input data increases, we use UMAP\cite{mcinnes2018umap} to reduce the dimensionality of SPECTER embeddings prior to clustering with HDBSCAN. 
We use multi-objective Bayesian hyperparameter tuning\cite{turner2021bayesian} for the UMAP-HDBSCAN pipeline to balance five evaluative criteria related to balancing inter- vs. intra-cluster density, number of clusters, and persistence of clusters over multiple runs of the algorithm. The successfully clustered papers are considered ``strong members'' of that cluster.


We refer to papers that cannot be confidently assigned by the clustering algorithm as ``weak members''. We assign each weak member to the cluster with which it has the highest semantic similarity. Downstream analyses are reported with and without inclusion of weak members. We consider this distinction because we suggest that weak members represent research which is significantly different (and potential truly innovative) relative to existing disciplines, and as such can help explain shifts in the trajectories of fields.\footnote{HDBSCAN refers to these non-confident assignments as noise; however, we expect these not to be noise in the traditional sense (e.g., an outlier or data worthy of discarding as it provides no analytical value) but instead to potentially add value as extremely novel research.}

\begin{figure}[htbp]
\centering
\includegraphics[width=\textwidth]{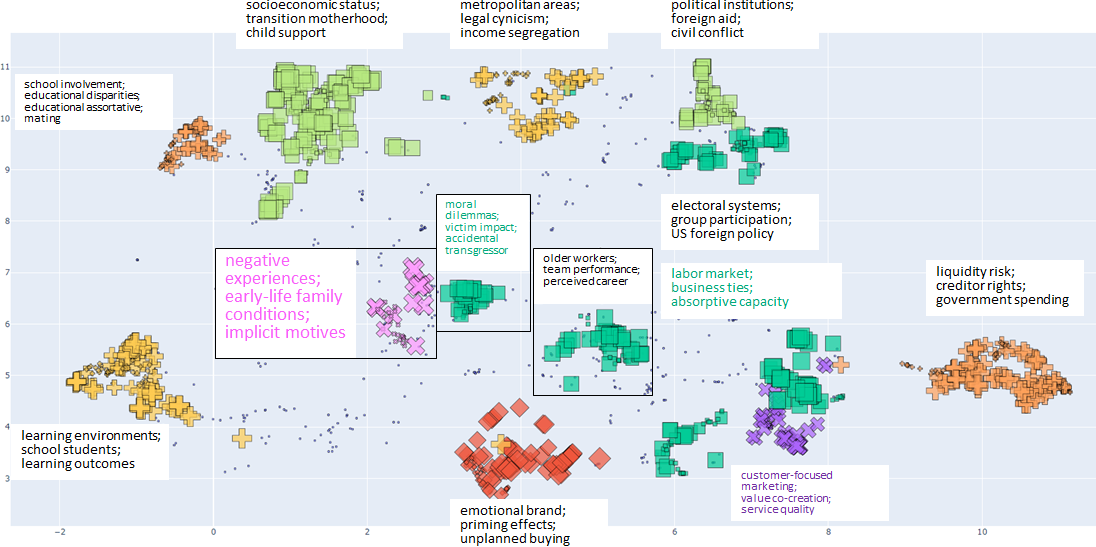}
\caption{UMAP projection of publications in the year 2011 colored by HDBSCAN-generated cluster labels with corresponding cluster-level keyphrases. Each cluster plotted here contains at least 2.5\% of total papers for the year and the size of each point is proportional to that publication's language-based interdisciplinarity score. Small blue points represent weak members. Note that most clusters shown are well-separated and not homogeneous in shape, suggesting that UMAP is doing a good job of dimensionally reducing the feature space in such a way that it is reasonably straightforward to partition and that a variable-density-based clustering algorithm, such as HDBSCAN, is well-suited to identifying clusters in such a dataset.} 
\label{fig:clusters}
\end{figure}

For each cluster, we generate representative keyphrases using a procedure similar to the KeyBERT library\cite{grootendorst2020keybert}, with modifications (e.g., more performant aggregation of embeddings from large numbers of documents belonging to the same cluster). Deriving keyphrases provides explanatory power for clusters and adds more nuanced understanding of the clusters than other commonly used approaches to grouping knowledge products, e.g., WoS categories. Clusters identified in our dataset for the year 2011 and their corresponding keyphrases are shown in Figure \ref{fig:clusters}. 
Our approach identifies a total of 371 clusters over the dataset, i.e., years 2011 through 2018.  

\subsection*{Citation graphs and communities}
\begin{figure}
    \centering
\includegraphics[width=0.9\textwidth]{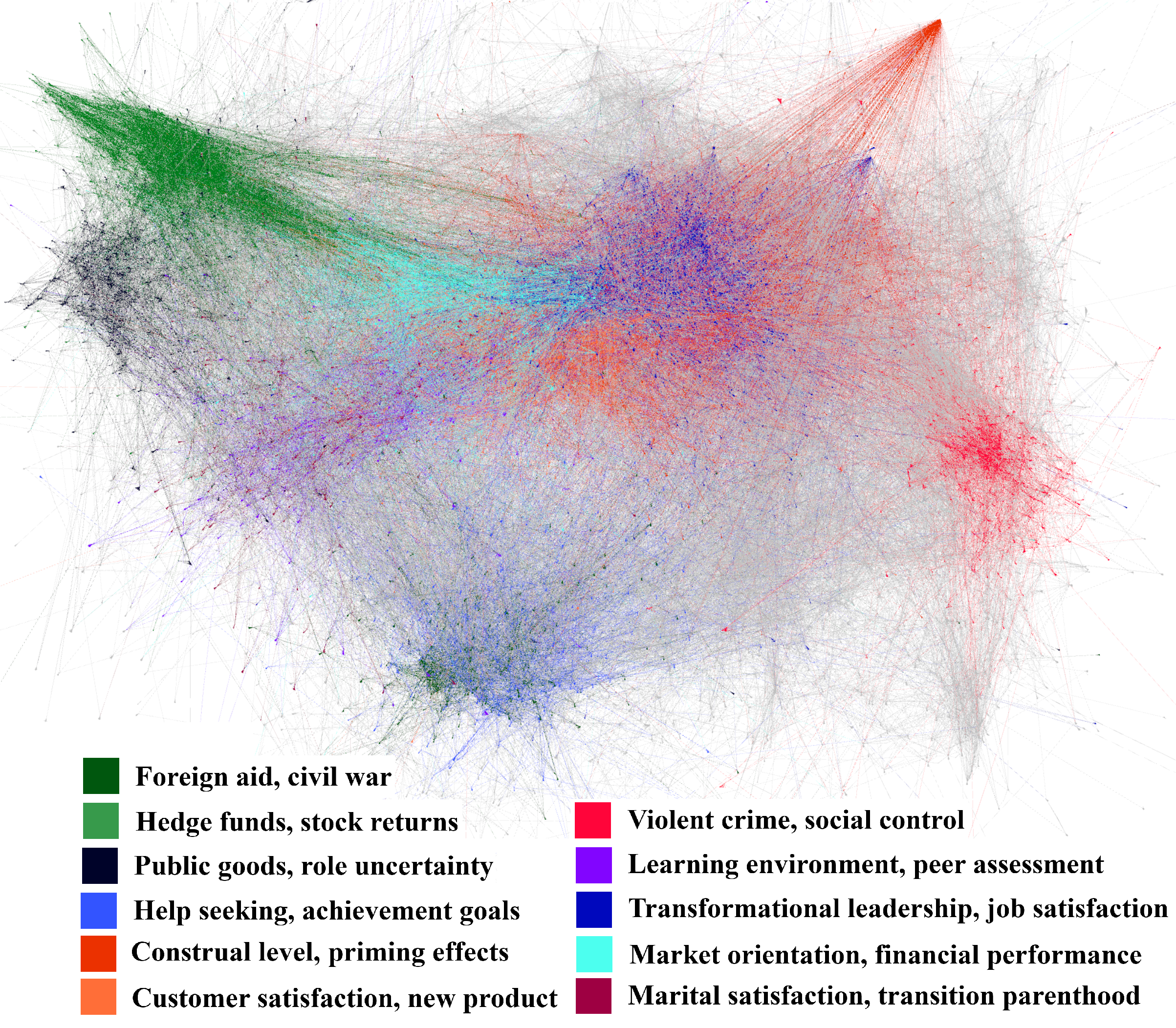}
\caption{An exemplary snapshot of the dense network and communities found by the Louvain community detection algorithm for the year 2011. Communities comprising less than 2.5\% of total papers for the year are colored grey. Note that a clear community structure can be observed for this graph-only approach much like it was for the language-only clustering presented earlier.}
\label{fig:communities}
\end{figure}

Per common practice, our citation-based analysis considers the citation network wherein nodes in the graph represent papers in our dataset and undirected edges represent citation relationships. We detect communities in this network using the Louvain community detection algorithm\cite{lu2015parallel}. 
Commonly-used, it maximizes modularity of the network, namely the expected value of inter- vs intra-community edges\cite{lancichinetti2009community}. Specifically, for a given time window/year of interest $t$ we consider the subgraph $G(t)$ containing only papers published in year $t$ and earlier, as well as their references.  This approach allows us to make predictions for past papers without fear that future papers citing them will cause information leakage into the dataset (e.g. a model trying to predict the evolution of an idea tied to a paper from 2017 should not have access to information about papers from 2018 citing it during model training). An example of the community structure discovered via the Louvain method is shown in Figure \ref{fig:communities}.

\subsection*{Quantifying interdisciplinarity}
\smallskip
\noindent \textbf{Language-based interdisciplinarity:} Our text-based interdisciplinarity (ID) metric scores each publication based on its soft clustering membership probabilities (i.e. the probability of a publication belonging to each possible cluster identified by standard or ``hard'' clustering), considering only strong member publications. 
It does so by assuming that one representation of interdisciplinarity is the diversity of language pulled from different fields. This metric is calculated using Equation \ref{eq:1} which considers the spread in its cluster assignment probabilities. Formally:
\begin{equation}\label{eq:1}
    \textnormal{ID}_{text} =\frac{N}{N-1}\left(1-P_{wm}-\max\left(P_{cluster}\right)\right)\left(1-\sigma_p\right)
\end{equation}
where $N$ is the total number of clusters in the dataset, $P_{cluster}$ is the probability of the paper belonging to a cluster, $P_{wm}$ is the probability of the paper being a weak member of any cluster, and $\sigma_p$ the standard deviation of $P_{cluster}$ over all clusters. 
This formulation is more intuitive when extreme cases are considered. For example, consider a corpus with 9 clusters for the year of interest. Consider a paper that sits very clearly within a single well-defined scientific discipline, i.e., $max(P_{cluster})=1$ for a single cluster (consequently, $P_{wm}=0$). The interdisciplinarity score for that paper would be $ID_{text}=0.0$. Alternatively, imagine a paper with membership probabilities that are equivalent for all clusters, with the same probability that it may be a weak member, i.e., $P_{wm}=P_{cluster,i}=0.1$ for $N=9$. This would result in $ID_{text}=0.9$, reflecting that the paper belongs to a wide array of disciplines/clusters equally, but also there is some chance that it may be a weak member -- which can also be interpreted as a global uncertainty in the membership probabilities -- thus keeping it from achieving a score of $1.0$. 

\medskip
\noindent \textbf{Citation-based interdisciplinarity:} We use betweenness centrality for each publication in the network as an interdisciplinarity metric, with higher centrality generally indicating higher interdisciplinarity, as has been done in previous literature\cite{rafols2007diversity}. 
As we do for community detection, we use time-windowed subgraphs for centrality measurement. 
Betweenness centrality is lightly modified for use as an ID metric, normalized on a [0,1] scale. For paper $i$ in publication year $t$:
\vspace{-0.2cm}
\begin{equation}\label{eq:2}
    \textnormal{ID}_{network} = centrality_{t,i}/max(\{centrality_{t}\})
\end{equation}
where $\{centrality_{t}\}$ is the set of all centrality values for papers published in calendar year $t$.

{
%
}

\subsection*{Text-based dynamic event modeling}

We identify and track critical knowledge evolution events borrowing from the literature tracking communities in dynamic social networks\cite{greene2010tracking}. Specifically, representative embeddings for each cluster are calculated using the element-wise mean of embeddings of the papers in each cluster, and clusters are compared across consecutive years by calculating the pairwise cosine similarity of the embeddings of each $[C_t, C_{t+1}]$ pair of clusters in years $t$ and $t+1$\cite{greene2010tracking}. We then link a cluster with its best-matching cluster(s) in the consecutive time step if the cosine similarity is above 0.95. 
We employ the following taxonomy\cite{greene2010tracking}:

\begin{itemize}
    \item A \emph{birth} event is identified at time $t$ when a cluster at time $t$ has no matching cluster(s) at time $t-1$.
    \item A \emph{death} event is identified at time $t$ when a cluster at time $t$ has no matching cluster(s) at time $t+1$. 
    \item Multiple clusters have \emph{merged} at time $t$ when one cluster at time $t$ matches to two or more clusters at time $t-1$.
    \item Multiple clusters have \emph{split} at time $t$ when one or more clusters at time $t$ match to a single cluster at time $t-1$.
    \item A \emph{continuation} event is observed when one cluster at time $t$ is matched to exactly one cluster at time $t+1$.
\end{itemize}

\begin{wrapfigure}{R}{0.48\textwidth}
\centering
\includegraphics[width=0.48\textwidth]{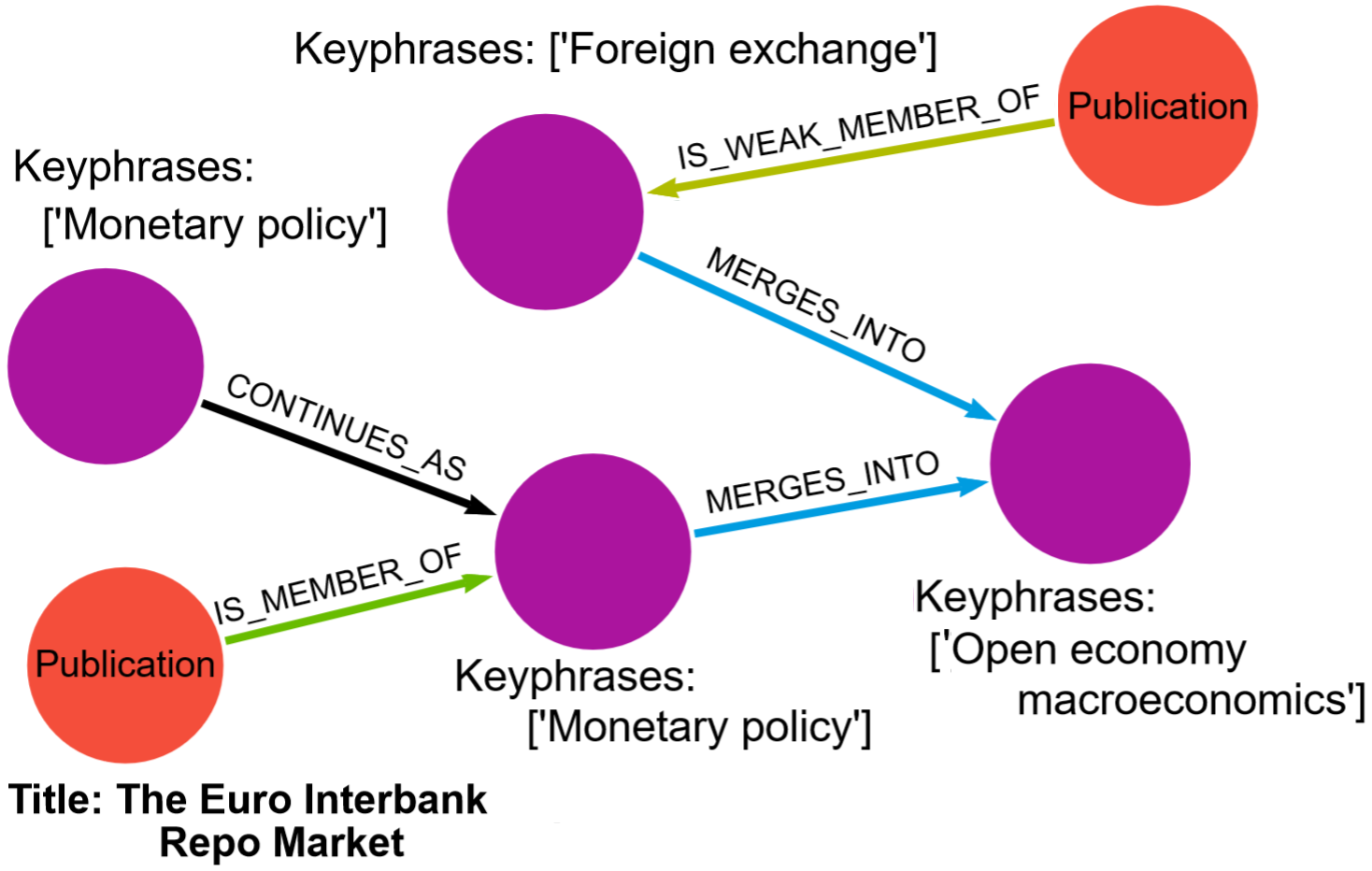}
\caption{Notional continuation and merge events showing weak (significantly different from existing clusters) and strong members (high confidence in its membership) of each cluster.}
\label{fig:EventSchematic}
\vspace{-0.5cm}
\end{wrapfigure}

\noindent We group these events into two types for subsequent analyses: (1) \emph{dynamic} -- split or merge and (2) \emph{stable} -- continuity or death.\footnote{Not only does treating splits and merges as a single class emerge from our metastability mental model but, given that they often co-occur, this treatment creates non-overlapping classes. We disregard birth events at present since a birth event has no preceding data from which to build a model and is unrelated to the concept of combinatorial innovation being described by metastability.} 
Figure \ref{fig:EventSchematic} gives a notional example of merge and continuation events. We note that events may occur in combination; e.g., a cluster may split into two, and those two clusters may simultaneously merge with two other clusters.



\subsection*{Event-tracking and prediction}\label{subsec:methods-explanation-prediction}

We hypothesize that interdisciplinarity scores and cluster size are indicators of metastability and therefore can be used to predict cluster evolution, i.e., \emph{dynamic} vs. \emph{stable} events, as an endogenous and target variable.
In particular, for each language cluster $C_t$ at time $t$, we use as exogenous model inputs: cluster-wise mean language-based interdisciplinarity score (which does include weak member papers);
mean citation-based interdisciplinarity score for weak and strong members, treated as separate features in order to discern if there is any difference in predictive power considering weak members; and number of weak and strong member papers in the cluster. 

To choose the most powerful features and test their predictive power (and thus value for further analyses), we use multinomial logistic regression and 
a Random Forest classifier with a binary target $\vec{y}$ representing if a dynamic event type (split or merge) is observed for a cluster at time $t+1$ as shown in Equation \ref{eqn:target-vector}.
\begin{equation}
\vec{y}=
\begin{bmatrix}
       split/merge \\
       continuation/death
\end{bmatrix}
\label{eqn:target-vector}
\end{equation}

\noindent We use the entire dataset with multinomial logistic regression for explanatory power. For the random forest, we use cluster events in the period 2011-17 for training and the year of 2018 for testing, resulting in roughly an 86\%/14\% train/test split by cluster count with 275 events for training (split/merge: 136; continuation/death: 139) and 43 testing events (split/merge: 21, continuation/death: 22). 
Using the above input features and event types in year $t+1$, we fit a random forest with 100 trees using the default hyperparameter values from the \code{scikit-learn} python library\cite{scikit-learn}. 

\section*{Results}
In the following, we first show that language and network frameworks capture different information by comparing the overlap between clusters identified using text and citation-based communities. We then further investigate the nature of the information provided by both frameworks by discussing how these representations, when considered together, not only serve to predict the evolution of disciplines and sub-fields but are equally important when doing so.

\subsection*{Comparing clusters and communities suggests valuable incomplete overlap}
\begin{wrapfigure}{r}{0.5\textwidth}
\centering
\includegraphics[width=0.5\textwidth]{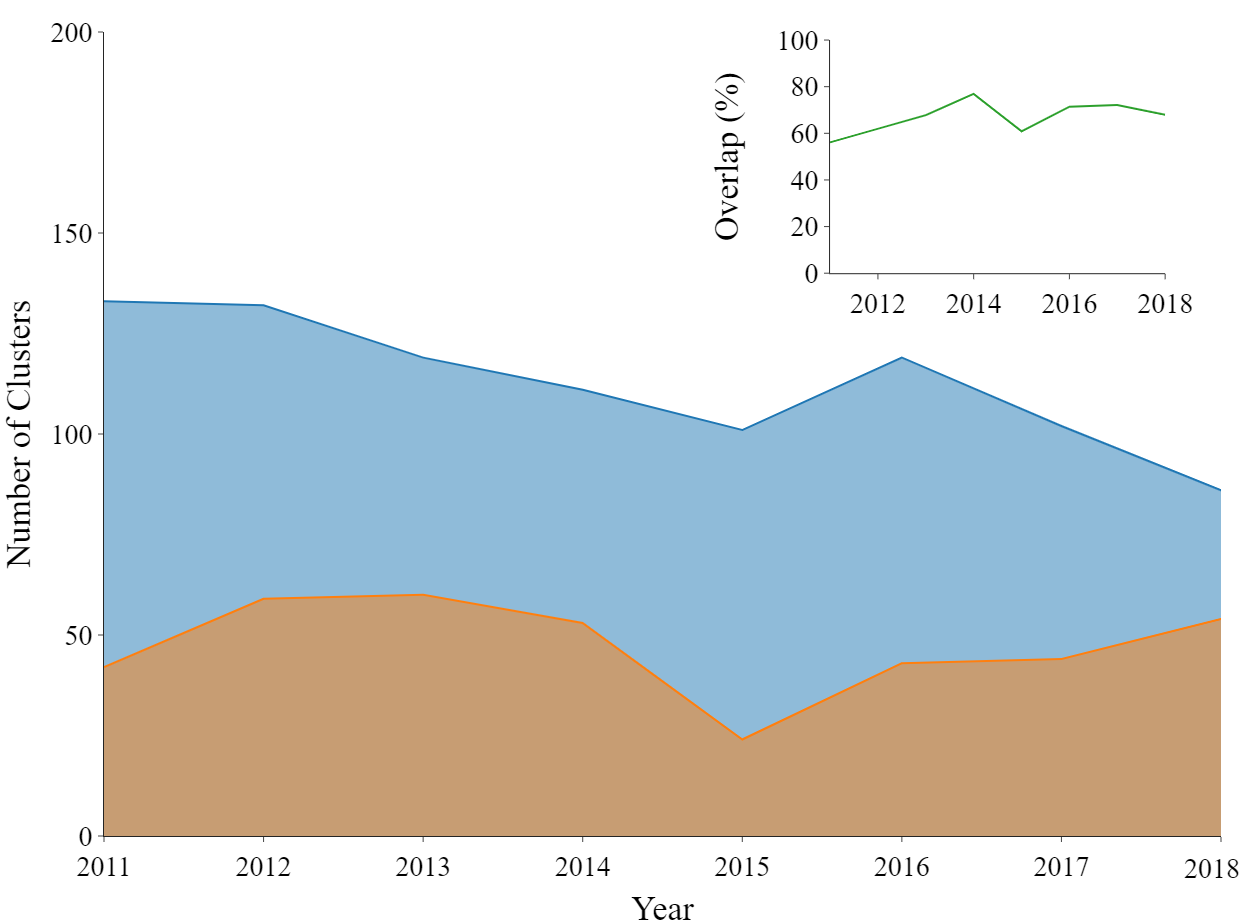}
\caption{Plot with number of clusters/communities identified by text-based (brown) and networks-based (blue) frameworks with inset plot showing percentage of language clusters associated with at least one network-derived community. Note that overlap values are consistently below 100\% but well above 0\%, suggesting unique and complementary insights added by each.}
\label{fig:overlap}
\vspace{-0.7cm}
\end{wrapfigure}

Figure \ref{fig:communities} gives a snapshot of network communities in 2011; comparison with Figure \ref{fig:clusters} illustrates differences in grouping across the two approaches. In general, the Louvain algorithm detects communities in the citation network at a finer resolution than our text-based clustering. For reference, Figure \ref{fig:overlap} shows the number of clusters and communities in our dataset, in addition to a measure of overlap between the two that we describe below. The number of network communities generally decreases over time, reflecting a more integrated citation graph emerging amongst the papers in our sample. 

As both our language- and citation-based frameworks are unsupervised, to compare them we need to identify clusters with one another across frameworks. For this, we measure pairwise Jaccard similarity between clusters and communities, effectively looking at the fraction of shared publications between every language cluster and every network community relative to their total number of member papers. If the similarity between a cluster and a community is above $0.1$ then we consider them similar. This threshold-based method (and the 0.1 threshold specifically) has been used in the literature for tracking clusters and communities over time \cite{greene2010tracking,asur2009event} and performs well across a variety of synthetic graphs. 
Going back to Figure \ref{fig:overlap}, the inset shows the percentage of language clusters with similar (Jaccard similarity $> 0.1$) network communities. It can be seen that while there is an overlap between the communities and clusters, the overlap is not complete, which suggests that each approach adds unique insight. 

\subsection*{Illustrating knowledge evolution events}

\begin{figure}[htbp]
\centering
\includegraphics[width=\textwidth]{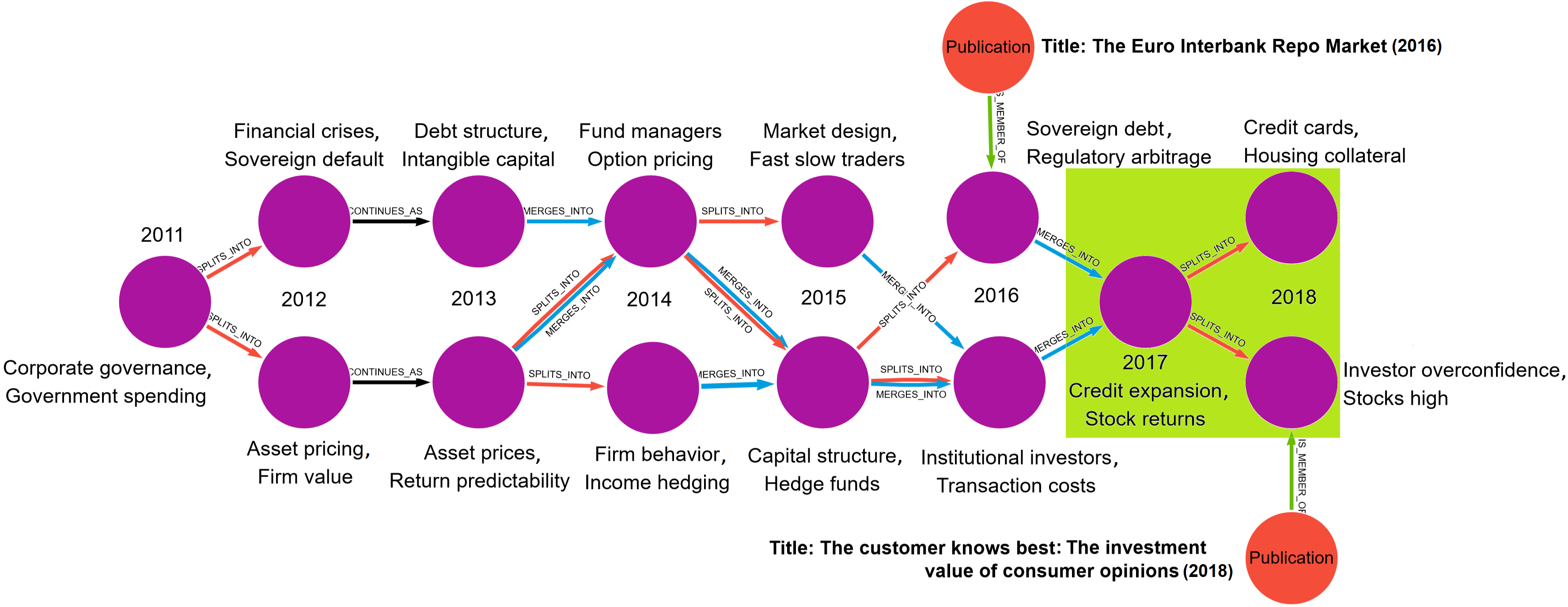}
\caption{Figure showing evolution of a set of language clusters from 2011 to 2018 (left to right) and keyphrases for each, along with two representative papers for two of the clusters. Note the marked change in focus between 2016 and 2018 evidenced by representative titles and cluster keyphrases. The split event for the 2017 cluster was successfully predicted by the random forest classifier described later (green box).}
\label{fig:example}
\end{figure}

To illustrate the types of knowledge events we identify and track in this work, let us consider an example from our dataset. 
Figure \ref{fig:example} shows the evolution of a full chain of language cluster evolutionary events over the period 2011 through 2018. Every cluster in this chain has 'Business and Finance' and 'Economics' as the most common WoS categories among member papers. In contrast, the keyphrases generated via our language clustering approach (shown in the Figure) reflect a much greater resolution, including phrases like ``income hedging'' and ``intangible capital''. This chain starts with a 2011 cluster that appears related to the (then recent) U.S. housing market crisis and Great Recession. There is a strong focus on work discussing corporate governance and government spending. This focus on organizational-level finance and economics mostly continues through 2017, with only a few deviations that are more focused on overall market trends. This is epitomized by the representative paper for one of the 2016 clusters, focused on European banking. Then something happens in 2018: topics appear to shift substantially from organizational/macroeconomic concepts to research focused on individual-level spending, finance, and decision-making, as can be seen both from the keyphrases representing those linguistic clusters, as well as from the representative 2018 paper focused on accounting for consumer behaviors in investing. It is interesting to note that this timing corresponds with Richard Thaler's 2017 Nobel Prize in Economics, awarded for contributions to behavioural economics.  


\subsection*{Knowledge evolution is significantly associated with interdisciplinarity and weak members}\label{subsec:mnlogit-results}

We use 
multinomial logistic regression with the mentioned endogenous and exogenous variables to evaluate how knowledge evolution may be explained through our interdisciplinarity scores, cluster size and network metrics. 
Per common practice, we insert a constant and a year variable to account for potential temporal effects. We attempt to explain whether or not clusters split or merge first, in order to evaluate the strength of associations between our hypothesized inputs and outputs.

Per Table \ref{tab:mnlogit-main-results}, we see significant positive associations between a cluster splitting or merging and the language interdisciplinarity score and network interdisciplinarity score with only certain associations (i.e., without weak members).\footnote{Following common best practice, we first conducted tests with all features, and, finding some insignificant, repeated with only significant features. See Supplementary Materials for details of this purposeful selection.}
We also see a positive association with the number of weak members associated with a cluster, and a negative association with the year.\footnote{Year was included per common practice to remove potential associations from time passing. Note that this model had a higher pseudo $R^2$ than a model without the year included. Future work should investigate any temporal associations through e.g., time series analyses.} Though all marginal effects are on the same order of magnitude, ranking by those effects, the language interdisciplinarity score is most important, followed by the number of weak members and the network score without weak members. Next, we further investigate this statistical relationship by testing the predictive power of a model trained on only a subset of these cluster data.

\begin{table}
\centering
\begin{tabular}{|l|r|r|r|r|} 
\hline
&\multicolumn{2}{r|}{\textbf{Model estimates}}&\multicolumn{2}{r|}{\textbf{Marginal effects}}\\
\hline
\textbf{Model Input (per cluster)} & \textbf{Coefficient} & \textbf{$P$} & \textbf{Effect} & \textbf{$P$}\\
\hline
Mean language ID score (strong members only) & 0.534 & 0.000 & 0.116 & 0.000\\
\hline
Number of weak members & 0.449 & 0.003 & 0.097 & 0.002\\
\hline
Mean network ID score (strong members only) & 0.292 & 0.030 & 0.063 & 0.025\\
\hline
Publication year & -0.372 & 0.007 & -0.081 & 0.005\\
\hline
Constant & -0.009 & 0.941 & &\\
\hline

\end{tabular}
\caption{Multinomial logistic regression results describing associations with split or merge ($1$) vs. continuation or death ($0$). Note significant positive associations with language score, network score, number of weak members, and a negative association for the year. All other features were not significant, and left out via purposeful selection for a more parsimonious model; see Supplemental Materials.}
\label{tab:mnlogit-main-results}
\end{table}

\subsection*{Validating our statistical result with predictive power - equal importance of interdisciplinarity scores}\label{subsec:randomforest-results}

\begin{table}
\centering
\begin{tabular}{|l|r|} 
\hline
\textbf{Model input feature} & \textbf{Gini Importance} \\
\hline
Mean language ID score (strong members only) & 0.336\\
\hline
Number of weak members & 0.234\\
\hline
Mean network ID score (strong members only) & 0.315\\
\hline
Publication year & 0.115\\
\hline
\end{tabular}
\caption{Random forest results on a held-out test set predicting the different types of cluster events a given cluster would experience in the next year, with the same features as in Table \ref{tab:mnlogit-main-results}. We achieved a micro-averaged $F_1 = 0.67$ on our held-out test set, with a class-specific $F_1 = 0.71$ for the class representing knowledge evolution (splits and merges). Per reported Gini feature importance of each independent variable, both interdisciplinarity scores are equally important, followed by number of weak members, then year. Note that the sort order of this table is identical to that of Table \ref{tab:mnlogit-main-results} to allow for more direct comparison of logistic regression coefficients to random forest feature importances.}
\label{tab:results-random-forest}
\end{table}

We have shown significant associations between knowledge splitting and merging, and interdisciplinarity and weak members. Here we go further by performing predictive modeling with a random forest classifier. Including only features shown to be statistically significant, we achieve a micro-averaged $F_1 = 0.67$ on our held-out test set, with $F_1 = 0.71$ on our class representing knowledge evolution (i.e. splitting or merging), a performance that is significantly better than random chance. Specifically, we present Table \ref{tab:results-random-forest}, which intuitively shows both interdisciplinarity scores to be equally important in achieving our predictive power. The number of weak members associated with a given cluster is next-most important, followed by the year variable. We validated against potential issues that can affect the Gini feature importance values from a random forest, specifically issues that arise when features exhibit multicollinearity and a bias towards numeric and high-cardinality categorical features \cite{strobl2007bias}. The first is not a problem in this case, as the high-correlation features were removed as a result of the logistic regression analysis discussed earlier. The second is expected to only be a minimal concern for this analysis, as the only non-numeric feature in this model is the publication year. Because this is a low-cardinality categorical feature, it may be the victim of a bias in the feature importances and, as a result, the year's true ranking in the feature importance table could be higher than is indicated. As this is not a critical change in the data for our analysis, correcting for this bias is beyond the scope of this work. Taken together, our results underscore the importance of including both the linguistic and network viewpoints of interdisciplinarity.

\section*{Conclusions and Future Work}

In this paper, we proposed a hybrid language- and network-based framework that uses state-of-the-art semantic embeddings and citation information to model metastability of ideas in order to identify dynamic events associated with the rise, fall, combination, and dispersion of topics in the scholarly corpus. We show that this hybrid approach is distinctly different from those based on linguistic or citation information alone. The approach we propose relies on a multi-dimensional view of interdisciplinarity as a predictor of scientific knowledge transitions.

Through both explanatory and predictive efforts, we show that language as well as network interdisciplinarity has positive effects on metastable knowledge combining and mixing. Interestingly, network interdisciplinarity of strong member papers is significantly predictive of these mixing events, even though the number of strong members is not. By contrast, even though weak members' network interdisciplinarity is not significantly predictive, more weak members are predictive of knowledge combining and mixing. This suggests that papers that do not cluster neatly are indicative of combinatorial innovation that is expressed as the knowledge mixing events discussed herein. As such, if one is interested in spurring broad interdisciplinarity, one should focus on encouraging more weakly-clustered research, regardless of its own network-derived interdisciplinarity. Future work should further investigate these relationships, in particular over longer time scales and on a more complete body of the scientific corpus. Additionally, a comparison of a few other useful and popular interdisciplinarity metrics is a natural extension of this work, to determine how well other established measures can predict the knowledge evolution events we have explored. A lack of consensus on the most useful interdisciplinarity metrics \cite{wang2020consistency} however makes this a challenge that must be tackled in later analyses.

This work also motivates and lays groundwork for new hybrid models that align multiple views of the literature ( e.g., linguistic, bibliometric) into unified modeling frameworks. Looking beyond traditional single-view approaches, such frameworks would be better suited to capture the richness of the scholarly record. This can be achieved through so-called graph machine learning modeling, that allows an integrated representation of a datum reflecting both its content (e.g. language in the case of a scientific paper) and its context within a network. Further, the work we describe here is mostly based on unsupervised learning. This is a necessity of the nature of this work, as there is no readily-available ground truth that is universally acknowledged to reflect the changing nature of scientific thought, disciplines, and sub-disciplines at a time scale reflective of how ideas mature and evolve. Future work should build benchmark datasets with which the metascience community can engage to evaluate and test these approaches more thoroughly than is currently possible. Possible proxy datasets that do exist at the moment include citation records -- the prediction of above-average citation growth, for example, could be another modeling task that is able to further determine the utility of the interdisciplinarity metrics presented in this paper.

\section*{Data Availability}
Parts of the data that support the findings of this study are available from Clarivate but restrictions apply to the availability of these data, which were used under license for the current study, and so are not publicly available. Data are however available from the authors upon reasonable request and with permission of Clarivate.
\section*{Code Availability}
The code used for data processing and model development for the current study is available at \url{https://github.com/QS-2/VESPID.git}.
\small
\bibliography{main}

\normalsize
\section*{Acknowledgements}

This research was supported by the National Center for Science and Engineering Statistics (NCSES) at the National Science Foundation through award 49100420C0030. The authors would like to thank Dr. Ashley Arigoni for her work on cluster comparison visualizations, as well as Mr. Joe Gorney and Mr. Alex Wade of Semantic Scholar for their aid in troubleshooting data engineering issues, and Dr. Ilya Rahkovsky of the Center for Security and Emerging Technology at Georgetown University and Dr. Phoebe Wong of Quantitative Scientific Solutions for their insights on the final analyses and drafts of this paper.

\section*{Author contributions statement}

D.G., S.R. conceived the experiment(s); S.K., D.R.M. and M.S. conducted the experiment(s).  All authors analyzed the results and reviewed the manuscript. 

\section*{Additional information}

The authors declare no competing interests.



\end{document}


\maketitle
\section*{Parameter selection with initial run of Logistic Regression}
We perform multinomial logistic regression to evaluate significance of endogenous variables to knowledge evolution, and to select parsimonious inputs for our predictive model. We use a simplified version of purposeful selection \cite{hosmer2013applied} to avoid overfitting and overdependence on observed data. Here we present detailed results of the initial run. Among the statistically insignificant variables, note that network interdisciplinarity's marginal effect was significant, so its presence was worth evaluating in a more parsimonious second model. Reported in our main results, this second model, confirmed the significance of network interdisciplinarity. We therefore dropped the other insignificant variables before our predictive model. We do note that pseudo-$R^2$ went down slightly (0.1025 vs. 0.1109), so future work should investigate the as-yet insignificant explanatory power that the removed exogenous variables provide as part of a wider effort to understand and explain knowledge metastability.

\begin{table}[!htbp]

\begin{tabular}  {|l|r|r|r|r|r|r|r|} 
\hline
&\multicolumn{6}{c|}{\textbf{Model estimates}}\\
\hline
\textbf{Model Input (per cluster)} & \textit{Coeff.} & \textit{Std. err.} & \textit{}{z} & \textit{$P>|z|$} & \textit{[0.025} & \textit{0.975]}\\
\hline
Cluster size (considering strong members only) & 0.010 & 0.006 & 1.638 & 0.101 & -0.002 & 0.021\\
\hline
Cluster size (considering weak members only) & 0.050 & 0.015 & 3.107 & 0.002 & 0.017 & 0.075\\
\hline
Mean language ID score & 2.704 & 1.189 & 2.274 & 0.023 & 0.374 & 5.035\\
\hline
Mean network ID score (considering strong members only) & 9.960 & 5.105 & 1.951 & 0.051 & -0.046 & 19.966\\
\hline
Mean network ID score (considering weak members only) & -3.801 & 4.878 & -0.779 & 0.436 & -13.361 & 5.759\\
\hline
Year & -0.179 & 0.072 & -2.495 & 0.013 & -0.320 & -0.038\\
\hline
Constant & 358.426 & 144.313 & 2.484 & 0.013 & 75.578 & 641.274\\
\hline
%
\hline
&\multicolumn{6}{c|}{\textbf{Marginal effects}}\\
\hline
\textbf{Model Input (per cluster)} & \textit{$dy/dx$} & \textit{Std. err.} & \textit{$z$} & \textit{$P>|z|$}&\textit{[0.025}&\textit{0.975]}\\
\hline
Cluster size (considering strong members only) & 0.002 & 0.001 & 1.660 & 0.097 & -0.000 & 0.005\\
\hline
Cluster size (considering weak members only) & 0.010 & 0.003 & 3.281   & 0.001 & 0.004 & 0.016\\
\hline
Mean language ID score & 0.580 & 0.247 & 2.351 & 0.019 & 0.096 & 1.063\\
\hline
Mean network ID score (considering strong members only) & 2.134 &  1.070 & 1.995 & 0.046 & 0.038 & 4.231\\
\hline
Mean network ID score (considering weak members only) & -0.815 & 1.042 & -0.782 & 0.434 & -2.856 & 1.227\\
\hline
Year & -0.039 & 0.015 & -2.589 & 0.010 & -0.067 & -0.009\\
\hline
\end{tabular}

\caption{Initial multinomial logistic regression results describing associations with knowledge evolution, binarized as split or merge ($1$) and continuation or death ($0$), and all our exogenous variables. Note significant positive associations with language score and number of weak members, plus a negative association for the year. Mean network interdisciplinarity score (considering strong members only) is statistically insignificant, as is the number of weak members. Mean network interdisciplinarity score (considering weak members only) is insignificant, yet it has a significant marginal effect. Per common practice, we purposefully re-ran our analysis discarding insignificant variables, to evaluate significance of network score among weak members and confirm our findings on language score, number of weak members, and year.} 
\label{tab:initial-mnlogit-results}
\end{table}

\bibliography{Supplemental}